\documentclass[11pt]{article} 

\usepackage{amsfonts}
\usepackage{amssymb}
\usepackage{amsmath}
\usepackage{graphicx}

\newcommand{\ket}[1]{|#1\rangle}

\newcommand{\bra}[1]{\langle #1|}

\newcommand{\op}[1]{\operatorname{#1}}
\def \qed {\hfill \rule{0.2cm}{0.2cm}\vspace{3mm}}
\newenvironment{mylist}[1]
	{\begin{list}{}{\setlength{\leftmargin}{#1}
	\setlength{\rightmargin}{0.0cm}\setlength{\labelsep}{1.3mm}
	\setlength{\labelwidth}{0.8cm}\setlength{\itemsep}{0.2cm}}}
	{\end{list}}

\newtheorem{theorem}{Theorem}

\newtheorem{cor}[theorem]{Corollary}
\newtheorem{prop}[theorem]{Proposition}
\newtheorem{definition}{Definition}

\begin{document}

\title{\Large\bf Succinct quantum proofs for properties of finite
groups\thanks{Research partially supported by Canada's NSERC.}
}

\author{
John Watrous\\
Department of Computer Science\\
University of Calgary\\
Calgary, Alberta, Canada\\
jwatrous@cpsc.ucalgary.ca}

\maketitle
\thispagestyle{empty}

\begin{abstract}\rm
In this paper we consider a quantum computational variant of
nondeterminism based on the notion of a {\em quantum proof}, which is a
quantum state that plays a role similar to a certificate in an NP-type proof.
Specifically, we consider quantum proofs for properties of
{\em black-box groups}, which are finite groups whose elements
are encoded as strings of a given length and whose group operations are
performed by a {\em group oracle}.
We prove that for an arbitrary group oracle there exist succinct
(polynomial-length) quantum proofs for the Group Non-Membership problem
that can be checked with small error in polynomial time on a quantum computer.
Classically this is impossible---it is proved that there exists a group
oracle relative to which this problem does not have succinct proofs that can
be checked classically with bounded error in polynomial time (i.e., the
problem is not in MA relative to the group oracle constructed).
By considering a certain subproblem of the Group Non-Membership problem we
obtain a simple proof that there exists an oracle relative to which BQP is not
contained in MA.
Finally, we show that quantum proofs for non-membership and classical proofs
for various other group properties can be combined to yield succinct
quantum proofs for other group properties not having succinct proofs in the
classical setting, such as verifying that a number divides the order of
a group and verifying that a group is not a simple group.
\end{abstract}


\section{Introduction}
\label{sec:introduction}

\noindent
There are several equivalent ways to view nondeterminism in the classical
setting that apparently yield inequivalent notions in the quantum setting.
Two such ways are as follows.

First, we may view a nondeterministic process as a probabilistic process,
and consider whether the resulting process has zero or nonzero probability of
success.
Along these lines, Adleman, DeMarrais, and Huang~\cite{AdlemanD+97} and Fenner,
Green, Homer, and Pruim~\cite{FennerG+99} have defined QNP to be
the class of languages $L$ for which there exist polynomial time quantum Turing
machines that accept with nonzero probability if and only if the input is in
$L$.
This class coincides with the counting class $\op{co-C}_=\op{P}$
\cite{FennerG+99, FortnowR99}.
This notion of quantum nondeterminism has also been investigated recently
in the context of communication complexity and query complexity by de~Wolf
\cite{deWolf00}.

Second, we way view nondeterminism as it relates to verification.
A common way to view NP is that NP is the class of languages consisting of
those strings for which there exist polynomial-length proofs of membership
that can be checked in polynomial time, and one may extend this viewpoint to
the quantum setting in several ways.
For instance, we may consider {\em quantum proofs} (or {\em quantum
certificates}), which are quantum states that certify membership of strings in
given languages, or we may consider ordinary (classical) certificates that are
checked by polynomial-time quantum computers.
In each case we may consider various constraints on the error allowed by the
quantum checking procedure.

In this paper, we investigate the second way of viewing nondeterminism in the
quantum setting.
We will restrict our attention to the case where certificates may be quantum
and the polynomial-time quantum verification procedure may operate with
(two-sided) bounded error.
Thus, this version of ``quantum NP'' represents the quantum generalization of
the class MA (based on the Arthur-Merlin games of
Babai~\cite{Babai85,BabaiM88}), and for this reason we will call the resulting
class QMA.
This notion of quantum nondeterminism was apparently first discussed by
Knill~\cite{Knill96}, and was later studied by Kitaev~\cite{Kitaev99}
(who instead referred to the class we call QMA as BQNP).
Kitaev proved $\op{QMA}\subseteq\op{P}^{\#P}$, and we claim that the technique
based on GapP functions used by Fortnow and Rogers~\cite{FortnowR99} to prove
$\op{BQP}\subseteq\op{PP}$ may be extended to prove $\op{QMA}\subseteq\op{PP}$
(this result was obtained jointly by A.~Kitaev and the present author).
One may also view QMA as a class that results by considering (two-sided error)
one-message quantum interactive proof systems
\cite{KitaevW00, Watrous99-qip-focs}, in which there is really no interaction
since only one message is sent.

Our main focus is on the power of QMA in the context of {\em black-box groups}.
Of particular interest to us is the Group Non-Membership problem, which may be
stated as follows:

\begin{center}
\underline{Group Non-Membership (GNM)}\\[2mm]
\begin{tabular}{ll}
Instance: & Group elements $g_1,\ldots,g_k$ and $h$ in some finite group $G$.\\
Question: & Is $h$ outside the group generated by $g_1,\ldots,g_k$
(i.e., is $h\not\in\langle g_1,\ldots,g_k\rangle$)?
\end{tabular}
\end{center}
The statement of this problem mentions neither the particular representation
of group elements used nor the underlying group or groups.
While it is interesting to consider this problem in the case that the group
elements are represented in some natural way (e.g., by invertible matrices
over a finite field), we will consider the case that group elements are
uniquely represented in some arbitrary way by strings, and that we have at our
disposal some oracle $B$ (known as a group oracle) that performs group
operations for us (with each operation requiring a single step).
In this setting, we assume nothing can be learned about group elements by
examining their representative strings except whether or not two elements are
distinct.
For each $n\in\mathbb{N}$ there will correspond a group consisting of some
subset of the length $n$ strings; this group will be denoted $B_n$ and is
called a {\em black-box} group.
Black-box groups were first considered by Babai and
Szemer\'edi~\cite{BabaiS84}, and have since been studied in several works
\cite{ArvindV97,Babai91,Babai92,Babai97,BabaiB99}.
Further details regarding black-box groups will be discussed in the next
section.

For a given group oracle $B$ we let $\op{GNM}(B)$ be the language consisting
of all positive instances of the Group Non-Membership problem relative to $B$.
By the Reachability Theorem of Babai and Szemer\'edi~\cite{BabaiS84}
it follows that $\op{GNM}(B)\in\op{co-NP}^B$ for any group oracle $B$.
Furthermore, Babai \cite{Babai91, Babai92} proved that
$\op{GNM}(B)\in\op{AM}^B$ for any group oracle $B$, while there exists choices
for the group oracle $B$ such that $\op{GNM}(B)\not\in\op{BPP}^B$ and
\mbox{$\op{GNM}(B)\not\in\op{NP}^B$}.
In Section~\ref{sec:oracle} we extend this result slightly by constructing a
group oracle $B$ such that \mbox{$\op{GNM}(B)\not\in\op{MA}^B$}.

In contrast to the fact that $\op{GNM}(B)\not\in\op{MA}^B$ for some choices of
the group oracle $B$, we prove that $\op{GNM}(B)\in\op{QMA}^B$ for any group
oracle $B$.
Thus, for any black-box group $G$ and elements $h,g_1\ldots,g_k\in G$, there
exists a polynomial-length quantum proof that $h$ is not in the
group generated by $g_1,\ldots,g_k$.
This fact is proved in Section~\ref{sec:non-membership}.
Naturally, a similar result holds in case group elements are represented in
any way that allows the group oracle to be replaced by a polynomial-time
computation, such as matrix groups over a finite field.
For such groups it is not known if GNM is in MA, although Babai~\cite{Babai92}
conjectures that in fact $\op{GNM}\in\op{NP}\cap\op{co-NP}$ in this restricted
case.
This conjecture is based on presently unproved conjectures relating to the
classification of finite simple groups.
A polynomial-time algorithm is known for permutation groups~\cite{Sims70}.

In certain limited cases it is possible to solve GNM in quantum polynomial
time without the help of a certificate, such as when $k=1$ in the statement
of the GNM problem.
The oracle $B$ we construct in Section~\ref{sec:oracle} in fact puts
$\op{GNM}(B)$ outside of $\op{MA}^B$ for this special case, and therefore
gives an oracle relative to which $\op{BQP}\not\subseteq\op{MA}$.
Bernstein and Vazirani \cite{BernsteinV93} claimed a stronger result
(specifically that there exists an oracle relative to which
$\op{EQP}\not\subseteq\op{MA}$), but the proof has not yet appeared.

Quantum proofs for group non-membership may be used to devise quantum
proofs for other group problems.
Several such problems, include the problem of testing whether a given number
divides the order of a group, testing that one group is a proper subgroup of
another, and testing that a given group is not a simple group, are
mentioned in Section~\ref{sec:other}.


\section{Definitions}
\label{sec:definitions}

\noindent
In this section we define the class QMA and discuss black-box groups in the
context of quantum circuits.
We assume the reader is familiar with the quantum circuit model, and with
basic notions from complexity theory and group theory.
For a detailed discussion of quantum circuits see Kitaev \cite{Kitaev97}.
(Readers not familiar with quantum computation may find the more introductory
papers of Berthiaume \cite{Berthiaume97} and Cleve \cite{Cleve99} helpful as
well.)
See, for example, Balc\'{a}zar, D\'{i}az, and Gabarr\'{o}
\cite{BalcazarG+88,BalcazarG+90} for background on complexity theory and, for
example, Isaacs~\cite{Isaacs94} for background on group theory.

Let us begin by making clear our assumptions regarding uniformity of quantum
circuits.
A family $\{Q_x\}$ of quantum circuits is said to be {\em polynomial-time
uniformly generated} if there exists a deterministic procedure that, on input
$x$, outputs a description of $Q_x$ and runs in time polynomial in $|x|$.
(For simplicity we assume all input strings are over the alphabet
$\Sigma = \{0,1\}$.)
It is assumed that the circuits in such a family are composed of gates in
some reasonable, universal, finite set of quantum gates (for instance, the
{\em standard basis} discussed by Kitaev \cite{Kitaev97} or the {\em Shor
basis} discussed by Boykin, et.~al.~\cite{BoykinM+99}).
In addition the circuits may include oracle gates as discussed below.
Furthermore, it is assumed that the size of any circuit in such a family is not
more than the length of that circuit's description (i.e., no compact
descriptions of large circuits are allowed), so that $Q_x$ must have size
polynomial in $|x|$.
To make matters simple when dealing with oracle gates below, we define the
size of a quantum circuit to be the number of gates in the circuit plus the
number of qubits upon which the circuit acts.

When we describe quantum circuits, we do so in a high-level manner that
may suggest that measurements are taking place at various times during the
circuit's computation; such measurements, however, do not occur and are
assumed to be simulated in the sense described by Aharonov, Kitaev, and
Nisan~\cite{AharonovK+98}.

For each circuit $Q_x$, some number of the qubits upon which $Q_x$ acts are
specified as {\em input qubits}, and all other qubits are {\em ancilla qubits}.
The input qubits are assumed to be initialized in some specified input state
$\ket{\psi}$, while all ancilla qubits are initialized to the $\ket{0}$ state.
One of the qubits is also specified as the {\em output qubit} and is assumed
to be observed after the circuit has been applied.
The probability that $Q_x$ accepts $\ket{\psi}$ is defined to be the
probability that an observation of the output qubit (in the
$\{\ket{0},\ket{1}\}$ basis) yields 1, given that the input qubits are
initially set to~$\ket{\psi}$.

We now define the class QMA as follows.
\begin{definition}\em
A language $A\subseteq\Sigma^{\ast}$ is in QMA if there exists a
polynomial-time uniformly generated family of quantum circuits
$\{Q_x\}_{x\in\Sigma^{\ast}}$ such that
(i)~if $x\in A$ then there exists a quantum state $\ket{\psi}$ such that
$\op{Pr}[\mbox{$Q_x$ accepts $\ket{\psi}$}] > 2/3$, and (ii)~if $x\not\in A$
then for all quantum states $\ket{\psi}$, $\op{Pr}[\mbox{$Q_x$ accepts
$\ket{\psi}$}] < 1/3$.
\end{definition}
Note that the circuit $Q_x$ does not take $x$ as an input, but rather the
procedure that produces the description of $Q_x$ takes $x$ as input---the
input $\ket{\psi}$
to a given circuit $Q_x$ corresponds to a quantum certificate that purportedly
proves the property that $x\in A$.
Information regarding $x$ may of course be ``hard-coded'' into $Q_x$, however,
which eliminates the need for inputting $x$.
It should be noted that the class QMA would not change if the definition was
such that there were just one circuit for each input length (rather than each
input), with each circuit taking $\ket{\psi}$ and $x$ as input (as would be the
case for the more standard notion of circuit uniformity).

Similar to classical bounded error classes, the bounds of 1/3 and 2/3 in the
definition of QMA may be replaced by $2^{-p(|x|)}$ and $1-2^{-p(|x|)}$,
respectively, for any polynomial $p$.
In the other direction, the bounds of 1/3 and 2/3 may be replaced by functions
$b(|x|)$ and $a(|x|)$, respectively, for $a,b:\mathbb{Z}^{+}\rightarrow[0,1]$
such that (i) $a$ and $b$ are computable in polynomial time, and
(ii)~\mbox{$a(|x|) - b(|x|)\geq 1/p(|x|)$} for some polynomial $p$.
In both cases, this follows from the fact that for any polynomial $q$
we may run $q(|x|)$ independent copies of a given verification procedure on
a ``compound certificate'' consisting of $q(|x|)$ certificates for the
independent copies, and make a decision to accept or reject depending on the
proportion of the individual copies that accept appropriately.
A simple analysis reveals that entanglement among the individual certificates
can yield no increase in the probability of acceptance as compared to the
situation in which the certificates are not entangled, and that the
probability of error is bounded by the tail of a binomial series as expected.

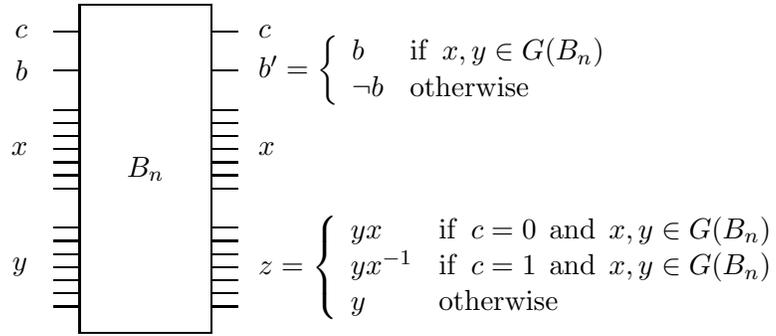
\begin{figure}[t]
\begin{center}
\begin{picture}(210,135)(80,50)
\setlength{\unitlength}{65000sp}%
\put(100,50){\framebox(50,125){$B_n$}}

\put(100,60){\line(-1,0){10}}
\put(100,65){\line(-1,0){10}}
\put(100,70){\line(-1,0){10}}
\put(100,75){\line(-1,0){10}}
\put(100,80){\line(-1,0){10}}
\put(100,85){\line(-1,0){10}}
\put(100,90){\line(-1,0){10}}

\put(100,105){\line(-1,0){10}}
\put(100,110){\line(-1,0){10}}
\put(100,115){\line(-1,0){10}}
\put(100,120){\line(-1,0){10}}
\put(100,125){\line(-1,0){10}}
\put(100,130){\line(-1,0){10}}
\put(100,135){\line(-1,0){10}}

\put(100,150){\line(-1,0){10}}

\put(100,165){\line(-1,0){10}}

\put(150,60){\line(1,0){10}}
\put(150,65){\line(1,0){10}}
\put(150,70){\line(1,0){10}}
\put(150,75){\line(1,0){10}}
\put(150,80){\line(1,0){10}}
\put(150,85){\line(1,0){10}}
\put(150,90){\line(1,0){10}}

\put(150,105){\line(1,0){10}}
\put(150,110){\line(1,0){10}}
\put(150,115){\line(1,0){10}}
\put(150,120){\line(1,0){10}}
\put(150,125){\line(1,0){10}}
\put(150,130){\line(1,0){10}}
\put(150,135){\line(1,0){10}}

\put(150,150){\line(1,0){10}}

\put(150,165){\line(1,0){10}}

\put(80,165){\makebox(0,0)[r]{$c$}}
\put(80,150){\makebox(0,0)[r]{$b$}}
\put(80,120){\makebox(0,0)[r]{$x$}}
\put(80,75){\makebox(0,0)[r]{$y$}}

\put(168,165){\makebox(0,0)[l]{$c$}}
\put(168,150){\makebox(0,0)[l]{$b'= \left\{\begin{array}{ll}b &
\op{if}\;x,y\in G(B_n)\\
\neg b & \op{otherwise}\end{array}\right.$}}
\put(168,120){\makebox(0,0)[l]{$x$}}
\put(168,75){\makebox(0,0)[l]{$z = \left\{\begin{array}{ll}
yx & \op{if}\;c = 0\;\op{and}\;x,y\in G(B_n)\\
yx^{-1} & \op{if}\;c = 1\;\op{and}\;x,y\in G(B_n)\\
y &  \op{otherwise}
\end{array}\right.$}}

\end{picture}
\end{center}
\caption{Reversible gate for a black-box group}
\label{fig:black-box}
\end{figure}

Next we will discuss black-box groups.
Here, we will consider a variation on black-box groups that is appropriate
for the quantum circuit model.
A group oracle $B$ is a family of bijections $\{B_n\}$ with
each member having the form $B_n:\Sigma^{2n+2}\rightarrow\Sigma^{2n+2}$ and
satisfying constraints to be discussed shortly.
We interpret the input and output of each $B_n$ as consisting of four parts:
a control bit, an error bit, and two $n$-bit strings representing group
elements.
This situation is pictured in Figure~\ref{fig:black-box}.
Associated with each $B_n$ is a group denoted $G(B_n)$ whose elements
form some subset of $\Sigma^n$ and whose group structure is determined
by the function $B_n$.
If $x,y\in G(B_n)$ then $yx = z$ for the unique value of $z$ that satisfies
$B(0,b,x,y) = (0,b,x,z)$ for each $b\in\Sigma$.
Similarly, if $x,y\in G(B_n)$ then $yx^{-1} = z$ for the unique value of $z$
that satisfies $B(1,b,x,y) = (1,b,x,z)$.
The first input bit (the control bit) thus determines whether $y$ is multiplied
(on the right) by $x$ or by $x^{-1}$.
Whenever we have $x\not\in G(B_n)$ or $y\not\in G(B_n)$, then it must be the
case that $B(c,b,x,y) = (c,\neg b,x,y)$ for each $b,c\in\Sigma$, i.e.,
the error bit $b$ is negated to indicate that the inputs were not valid group
elements.
Naturally, the constraint that must be obeyed by each $B_n$ in order for
$B = \{B_n\}$ to be considered a group oracle is that there must exist a
family of underlying groups $\{G_n\}$ along with encodings $\{f_n\}$ (each
$f_n:G_n\rightarrow\Sigma^n$ one-to-one and satisfying $f_n(G_n) = G(B_n)$)
that yields the above structure.
Each group $G(B_n)$, and more generally any subgroup of $G(B_n)$ given by
a list of generators, is known as a black-box group.

For a given group oracle $B$ each $B_n$ is invertible, and may therefore be
viewed as a $(2n+2)$-qubit quantum gate as suggested by
Figure~\ref{fig:black-box}.
When we say that a polynomial-time uniformly generated family of quantum
circuits has access to group oracle $B$, we mean that the circuits in the
family may, in addition to the standard gates mentioned previously, be
composed of any of the gates in the collection $\{B_n\}$.
Note that any quantum circuit containing a $B_n$ gate must have size
$\Omega(n)$.


\section{Verification of non-membership}
\label{sec:non-membership}

\noindent
In this section we prove that the Group Non-Membership problem is in QMA
for an arbitrary group oracle $B$.
Before giving the technical proof, we will discuss informally the basic idea
of the proof.

Suppose group elements $g_1,\ldots,g_k$ and $h$ are given, 
and let us write $H = \langle g_1,\ldots,g_k\rangle$.
Consider the state $|H|^{-1/2}\sum_{g\in H}\ket{g}$,
and assume that this state is contained in a quantum register $\mathbf R$.
In general, given any finite set $A$ we will let $\ket{A}$ denote the
state $|A|^{-1/2}\sum_{a\in A}\ket{a}$, so that we may say that
$\mathbf R$ is in state $\ket{H}$.
In addition let $\mathbf B$ be a register consisting of a single qubit, and
suppose $\mathbf B$ is initialized to state $(\ket{0}+\ket{1})/\sqrt{2}$.
Assuming we have a gate that performs group operations as discussed in the
previous section, we may built a quantum circuit acting on $\mathbf R$ and
$\mathbf B$ that effectively acts as a controlled-multiply-by-$h$ operation on
$\mathbf R$, where $\mathbf B$ is the control.
If this operation is performed, we may express the resulting state of the pair
$(\mathbf B,\mathbf R)$ as $(\ket{0}\ket{H} + \ket{1}\ket{Hh})/\sqrt{2}$.
Now perform a Hadamard transform on $\mathbf B$ to yield the state
\[
\frac{1}{2}\ket{0}(\ket{H}+\ket{Hh})+\frac{1}{2}\ket{1}(\ket{H}-\ket{Hh}).
\]
At this point, observing $\mathbf B$ in the $\{\ket{0},\ket{1}\}$ basis yields
1 with probability $p = \|(\ket{H}-\ket{Hh})/2\|^2$.
In case $h\in H$ we have $\ket{H}=\ket{Hh}$, and so $p=0$; in case $h\not\in H$
we have that $\ket{H}$ and $\ket{Hh}$ are orthogonal, and so $p = 1/2$.
Thus, given several copies of the state $\ket{H}$ one may determine with
very high probability whether or not $h\in H$.

Unfortunately, the state $\ket{H}$ may be difficult to construct in some
cases, but it may be given as a quantum certificate.
Naturally we may not assume that a given certificate $\ket{\psi}$ coincides
with $\ket{H}$, so this must be verified before the above test is performed.
In fact, it is not necessary to check that $\ket{\psi} = \ket{H}$, but only
that $\ket{\psi}$ is invariant under right multiplication by elements of $H$.
Our technique to do this is as follows.
Consider a (classical) randomized procedure for generating elements of $H$
uniformly (for now we assume this is possible without error---we will
take errors into account in the proof below).
We may modify such a probabilistic process to make it quantum by
simulating the act of choosing any random number in some given range
$\{0,\ldots,N-1\}$ by using a quantum transformation $Q_N$ satisfying
$Q_N\ket{0} = N^{-1/2}\sum_{a = 0}^{N-1}\ket{a}$, and simulating the
entire process reversibly.
(To do this, assume all random choices are made first, and that the remaining
part of the process is deterministic and hence can be simulated reversibly.)
Let $F$ denote the resulting quantum transformation.
It will not be the case that $F$ produces $\ket{H}$, but rather we will have
\[
F:\ket{0}\mapsto\frac{1}{\sqrt{|H|}}\sum_{g\in H}\ket{g}\ket{\op{garbage}(g)}
\]
for $\ket{\op{garbage}(g)}$ denoting some arbitrary unit vector representing
whatever is left over from this process (for instance, copies of the
simulated random numbers yielding the random choice of $g$ in superposition).
Now, to check that the state contained in $\mathbf R$, which purportedly
contains $\ket{H}$, is invariant under right multiplication by elements
of $H$, we do the following: (i) apply $F$ to some register $\mathbf S$
that is initially in the state $\ket{0}$, (ii) multiply (on the right) the
contents of $\mathbf R$ by the ``random'' group element contained in
$\mathbf S$, (iii) apply $F^{\dagger}$ to $\mathbf S$, and (iv) observe
$\mathbf S$.
If $\mathbf R$ was invariant under multiplication by elements of $H$,
then $\mathbf S$ will revert back to state $\ket{0}$ with certainty, while
if not there will be some probability that the observation of $\mathbf S$
yields some other result (indicating that this certificate should be rejected).
Under the assumption that the observation of $\mathbf S$ does yield 0,
however, the state of $\mathbf R$ will in fact be changed (by quantum magic!)
to one that is invariant under right multiplication by elements in $H$.
At this point, $\mathbf R$ will be suitable for the first test that determines
whether $h\in H$.

Before proceeding to the formal proof, we mention the following theorem due to
Babai~\cite{Babai91} that will be used in the proof.
The theorem essentially states that elements in a given black-box group can be
randomly generated in such a way that the resulting distribution is very close
to uniform.

\begin{theorem}[Babai]
\label{theorem:uniform}
For any group oracle $B$ there exists a randomized procedure
$\mathcal{P}$ acting as follows.
On input $g_1,\ldots,g_k\in G(B_n)$ and $\epsilon>0$, $\mathcal{P}$ outputs an
element of $H=\langle g_1,\ldots,g_k\rangle$ in time polynomial in
$n + \log 1/\epsilon$ such that each $g\in H$ is output with probability in
the range $(1/|H|-\epsilon,1/|H|+\epsilon)$.
\end{theorem}
This is in fact a weaker result than the one proved by Babai, but it is
sufficient for our needs.

Now we are prepared to state and prove the main result of this section.

\begin{theorem}
\label{theorem:non-membership}
$\op{GNM}(B)\in\op{QMA}^{B}$ for any group oracle $B$.
\end{theorem}

\noindent
{\bf Proof.}
As above, given any set $A$, we write $\ket{A}$ to denote the uniform
superposition over elements of $A$, i.e.,
$\ket{A} = |A|^{-1/2}\sum_{a\in A}\ket{a}$.
Let $g_1,\ldots,g_k$ and $h$ denote input group elements of length $n$, let
$H = \langle g_1,\ldots,g_k\rangle$, and consider the procedure described in
Figure~\ref{fig:arthur_non-membership}.
\begin{figure}[!ht]
\hrulefill
\begin{mylist}{0mm}
\item Assume register $\mathbf R$ contains the quantum certificate,
and all other registers are initialized to $\ket{0}$.

\item Let $F$ be a transformation such that
\[
F:\ket{0}\mapsto\sum_{g\in H}\alpha_g\ket{g}\ket{\op{garbage}(g)},
\]
where $|\alpha_g|^2\in\left(1/|H|-2^{-2n},1/|H|+2^{-2n}\right)$ for each
$g\in H$, and $\ket{\op{garbage}(g)}$ denotes some arbitrary unit vector that
depends on $g$.
The fact that transformation $F$ can be performed in by
polynomial-time uniform quantum circuits follows from
Theorem~\ref{theorem:uniform}, as described previously.

\vspace{2mm}
\item {\bf Step 1:}
\vspace{2mm}

Using the group oracle, check that $\mathbf R$ contains a valid
element of $G(B_n)$.  \underline{Reject} if this is not the case.

Apply transformation $F$ to register $\mathbf S$.

Using the group oracle, multiply the contents of register $\mathbf R$
by the group element contained in $\mathbf S$.

Apply transformation $F^{\dagger}$ to $\mathbf S$.
If $\mathbf S$ does not contain $0$, then \underline{reject}.
Otherwise proceed to step 2.

\vspace{2mm}
\item {\bf Step 2:}
\vspace{2mm}

Apply Hadamard transform to an initialized register $\mathbf B$ (i.e.,
set register $\mathbf B$ to state $(\ket{0} + \ket{1})/\sqrt{2}$).

Using the group oracle, perform a controlled-multiply-by-$h$ operation on
register $\mathbf R$, where $\mathbf B$ is the control bit.
(Specifically, this operation has the effect of multiplying the contents of
register $\mathbf R$ on the right by $h$ if $\mathbf B$ has value 1, and has
no effect if $\mathbf B$ has value 0.)

Perform a Hadamard transform on $\mathbf B$, and \underline{reject} if
$\mathbf B$ contains 0.

If the computation has not rejected thus far, then \underline{accept}.

\end{mylist}

\hrulefill
\caption{Quantum verification procedure for Group Non-Membership.}
\label{fig:arthur_non-membership}
\end{figure}

Assume first that $h\not\in H$.
In this case we must prove that there exists a certificate $\ket{\psi}$ causing
the procedure to accept with high probability.
The certificate will be $\ket{H}$.
The verification procedure first performs transformation $F$ on $\mathbf S$,
which was initialized to $\ket{0}$ at the start of the procedure.
The state of the pair of registers $(\mathbf R,\mathbf S)$ is now
\begin{equation}
\label{eq:step1}
\ket{H}\sum_{g\in H}\alpha_g(\ket{g}\ket{\op{garbage}(g)}).
\end{equation}
The contents of register $\mathbf R$ is multiplied by the group element
contained in $\mathbf S$, which has no effect on the state in (\ref{eq:step1})
following from the fact that $\ket{H}$ is invariant under multiplication by
any element $g\in H$.
Now the inverse of transformation $F$ is applied, which returns $\mathbf S$ to
the state $\ket{0}$ with certainty.
The probability that the verification procedure rejects in step 1 is therefore
0.
Now step 2 is performed.
After preparing register $\mathbf B$ and performing the
controlled-multiply-by-$h$ operation, the state of the pair
$(\mathbf B,\mathbf R)$ is
$(\ket{0}\ket{H} + \ket{1}\ket{Hh})/\sqrt{2}$.
A Hadamard transform is performed on $\mathbf B$, producing the state
\[
\frac{1}{2}\ket{0}(\ket{H}+\ket{Hh})+ 
\frac{1}{2}\ket{1}(\ket{H}-\ket{Hh}).
\]
Under the assumption $h\not\in H$, we have that $\ket{H}$ and $\ket{Hh}$
are orthogonal, and consequently the probability of acceptance is
$\left\|(\ket{H} - \ket{Hh})/2\right\|^2 = 1/2$.

Now suppose $h\in H$ and let $\ket{\psi}$ denote the initial state of register
$\mathbf R$.
In this case our goal is to bound the probability of acceptance.
Let us write
\[
\ket{\psi} = \sum_{x\in G(B_n)}\beta_x\ket{x} + \ket{\gamma}
\]
for $\ket{\gamma}\in\op{span}\{\ket{x}\,:\,x\not\in G(B_n)\}$ denoting
the ``invalid'' portion of $\ket{\psi}$.
The verification procedure first checks that $\mathbf R$ contains a
superposition over valid elements of $G(B_n)$, which has the effect of
projecting the state of $\mathbf R$ to $\sum_{x\in G(B_n)}\beta_x\ket{x}$
(renormalized) in case this test does not result in rejection.
As we are interested in bounding the overall (unconditional) probability of
accepting, however, we need not renormalize this state.
Transformation $F$ is performed on $\mathbf S$, and the group element contained
in $\mathbf S$ is multiplied to the contents of $\mathbf R$, producing state
\[
\sum_{x\in G(B_n)}\sum_{g\in H}\alpha_g \beta_x\ket{xg}\ket{g}
\ket{\op{garbage}(g)}
\]
in registers $(\mathbf R,\mathbf S)$.
Now $F^{\dagger}$ is applied to $\mathbf S$ and the verification procedure
rejects if $\mathbf S$ has not been returned to it's initial 0 value.
Under the assumption that an observation of $\mathbf S$ reveals 0 (which is
necessary if the procedure accepts), the state of register $\mathbf R$ becomes
\[
\sum_{x\in G(B_n)}\sum_{g\in H}\alpha_g \beta_x\ket{xg}
\bra{0}F^{\dagger}(\ket{g}\ket{\op{garbage}(g)})
\:=\: \sum_{x\in G(B_n)}\sum_{g\in H}|\alpha_g|^2 \beta_x\ket{xg}
\]
(where again we do not renormalize in order to calculate the unconditional
probability of acceptance).
Now step 2 is performed.
After the controlled-multiply-by-$h$ and Hadamard operations have been
performed, the state of the pair $(\mathbf B,\mathbf R)$ will be
\[
\frac{1}{2}\ket{0}\sum_{x\in G(B_n)}\sum_{g\in H}
\left(|\alpha_g|^2 \beta_x\ket{xg} + |\alpha_g|^2 \beta_x\ket{xgh}\right)
+ \frac{1}{2}\ket{1}\sum_{x\in G(B_n)}\sum_{g\in H}
\left(|\alpha_g|^2 \beta_x\ket{xg} -|\alpha_g|^2 \beta_x\ket{xgh}\right).
\]
The probability of acceptance is therefore
\begin{equation}
\label{eq:prob_acc1}
\frac{1}{4}\left\|
\sum_{x\in G(B_n)}\sum_{g\in H}
\left(|\alpha_g|^2\beta_x\ket{xg}-|\alpha_g|^2
\beta_x\ket{xgh}\right)\right\|^2.
\end{equation}
Under the assumption that $h\in H$, we have that $xgh$ and $xg$ range over
the same set as $g$ ranges over $H$.
Thus we may rewrite (\ref{eq:prob_acc1}) as
\begin{equation}
\label{eq:prob_acc2}
\frac{1}{4}\left\|\sum_{x\in G(B_n)}\sum_{g\in H}\beta_x
\left(|\alpha_g|^2-|\alpha_{gh^{-1}}|^2\right)\ket{xg}\right\|^2.
\end{equation}
By the triangle inequality, we see that (\ref{eq:prob_acc2}) is at most
\[
\frac{1}{4}\left(\sum_{g\in H}\left(|\alpha_g|^2 -
|\alpha_{gh^{-1}}|^2\right)
\left\|\sum_{x\in G(B_n)}\beta_x\ket{xg}\right\|\right)^2 \leq 2^{-2n}.
\]
Thus we have that the verification procedure accepts with exponentially small
probability.

The definition of QMA requires that positive instances be accepted with
probability at least 2/3 and negative instances to be accepted with
probability at most 1/3.
Thus, we must address the fact that although our verification procedure
accepts with exponentially small probability for all certificates on negative
instances, the probability of acceptance is only guaranteed to be 1/2 for
positive instances.
As discussed in Section~\ref{sec:definitions}, this may be remedied by
running several copies of the verification procedure in parallel and
deciding to accept or reject depending on the number of parallel
executions that accept.
In the present case we may achieve exponentially small probability of
error by running a polynomial number of copies of the above verification
procedure on a compound certificate and accepting if and only if at least
one of the copies accepts.
\qed


\section{Oracle separations}
\label{sec:oracle}

\noindent
In this section we discuss oracle separations regarding MA, QMA, and BQP.
First, we prove that there exists a group oracle $B$ relative to which the
Group Non-Membership problem is not contained in $\op{MA}$, and thus
$\op{MA}^B\subsetneq\op{QMA}^{B}$.
Our proof follows the same general ideas used by Babai~\cite{Babai91,Babai92}
to prove $\op{GNM}\not\in\op{NP}$ and $\op{GNM}\not\in\op{BPP}$ for some
group oracles.
We then identify a restricted version of the Group Non-Membership problem,
which we call the 2-Element Group Non-Membership problem, that in fact is
contained in BQP but still lies outside of MA relative to the group oracle $B$.
Thus we have an oracle separating BQP and MA.
A stronger result was claimed by Bernstein and Vazirani \cite{BernsteinV93},
but their proof has not yet appeared---they claimed the existence of an
oracle relative to which $\op{EQP}$ is not contained in $\op{MA}$.

The oracle separations we prove rely on a strong amplification property
possessed by MA, which is that the probability of error can be made much
smaller than the reciprocal of the number of possible certificates for each
input length.
With this in mind, we take the following as our definition of $\op{MA}^B$:
\begin{definition}\em
\label{def:MA}
For a given group oracle $B$, a language $A$ is in $\op{MA}^B$ if there exists
a predicate $R$, computable in polynomial time by a deterministic Turing
machine with access to the group oracle $B$, and polynomials $q$ and $r$, such
that for every $x\in\Sigma^{\ast}$ we have:

\begin{mylist}{\parindent}
\item If $x\in A$, then there exists $y\in\Sigma^{q(|x|)}$ such that
\[
\left|\left\{\left.z\in\Sigma^{r(|x|)}\right|R(x,y,z)=1\right\}\right|
\:=\: 2^{r(|x|)}.
\]

\item If $x\not\in A$, then for all $y\in\Sigma^{q(|x|)}$,
\[
\left|\left\{\left.z\in\Sigma^{r(|x|)}\,\right|\,R(x,y,z) = 1\right\}\right|
\: < \: 2^{-2q(|x|)}2^{r(|x|)}.
\]
\end{mylist}
\end{definition}
This definition also includes the fact that the error can be made one-sided
without changing the resulting class (see, for instance,
Zachos~\cite{Zachos88})---a property that we do not know holds for QMA.
This fact is not essential in our proof, but has the advantage of simplifying 
our analysis.

\begin{theorem}
\label{theorem:oracle}
There exists a group oracle $B$ for which we have
$\op{GNM}(B)\not\in\op{MA}^B$.
\end{theorem}
\noindent
{\bf Proof.}
For each $n\geq 4$, let $p(n)$ be a prime number satisfying
$2^{n-2}<p(n)^2<2^n$.
Existence of such a sequence of primes follows from Bertrand's Postulate,
first proved by Chebyshev (see, for instance, Rosser and Schoenfeld
\cite{RosserS62}).
Let $[p(n)^2]$ denote the set $\{1,\ldots,p(n)^2\}$, and for fixed $n$
identify each element of $[p(n)^2]$ with its representation as an $n$-bit
string in binary.
Let $\mathcal{F}(n)$ denote the set of one-to-one functions of the form
$f:[p(n)^2]\rightarrow \mathbb{Z}_{p(n)}\times\mathbb{Z}_{p(n)}$, and define
\begin{eqnarray*}
\mathcal{F}_1(n) & = & \left\{f\in\mathcal{F}(n)\,|\,f(1) = (1,0)\;
\mbox{and}\;f(2) = (0,1)\right\},\\
\mathcal{F}_0(n) & = & \{f\in\mathcal{F}(n)\,|\,f(1) = (1,0)\;\mbox{and}\;
f(2) = (a,0) \mbox{for some $a\in\{2,\ldots,p(n)-1\}$}\}.
\end{eqnarray*}
We have $|\mathcal{F}_0(n)| = (p(n)-2)|\mathcal{F}_1(n)|$.
Associated with each $f\in\mathcal{F}(n)$ is a black-box group
isomorphic to \mbox{$\mathbb{Z}_{p(n)}\times\mathbb{Z}_{p(n)}$} that labels
each $(\alpha,\beta)\in\mathbb{Z}_{p(n)}\times\mathbb{Z}_{p(n)}$ with the
$n$-bit string $f^{-1}(\alpha,\beta)$.
When $n$ is fixed, or understood from context, we will simply write $p$,
$\mathcal{F}_0$, $\mathcal{F}_1$, etc., to mean $p(n)$, $\mathcal{F}_0(n)$,
$\mathcal{F}_1(n)$, etc.

We will restrict our attention to the case where the input to the GNM problem
consists of the pair of $n$-bit strings representing labels $1$ and $2$
in binary for some $n$---we will write this pair as $(1,2)_n$ in order to
stress the dependence on $n$.
Furthermore, we also restrict our attention to the case that the group oracle
is associated with some $f\in\mathcal{F}_1(n)\cup\mathcal{F}_0(n)$ for each
$n$ as described previously.
For fixed $n$, if the group in question is associated with
$f\in\mathcal{F}_1$, then $f(2)\not\in\langle f(1)\rangle$, and so $(1,2)_n$
is a positive instance of GNM.
If the group is associated with $f\in\mathcal{F}_0$, then
$f(2)\in\langle f(1)\rangle$, and so $(1,2)_n$ is a negative instance of GNM.

Below we will diagonalize over all polynomial time oracle Turing machines
in order to prove the existence of $B$ as in the statement of the theorem.
First, let us consider an arbitrary polynomial-time deterministic oracle Turing
machine $M$, and let $q$, $r$, and $t$ be strictly increasing polynomials such
that the following holds: for any $x\in\Sigma^{\ast}$, $y\in\Sigma^{q(|x|)}$,
and $z\in\Sigma^{r(|x|)}$, $M$ runs in time $t(|x|)$ on input $(x,y,z)$ and
any group oracle $B$.
(Here, $x$, $y$, and $z$ are as in the definition of MA, i.e., $x$ corresponds
to the input, $y$ is a certificate, and $z$ is treated as a sequence of random
bits.)
As mentioned above, we are interested in the case where $x = (1,2)_n$ for some
$n$.
Write $m = |x|$ for such a choice of $x$, and for simplicity assume our
encoding of pairs of strings is such that $2n \leq m \leq 4n$.
At this point we will fix $n$ sufficiently large such that $8t(4n)^2<2^{n/2}$
(and thus $t(m)^2/p(n)<1/4$).
Let $B$ be an arbitrary group oracle, and for any $f\in\mathcal{F}$ let us
write $B_f$ to denote the new group oracle obtained by changing the behavior
of $B$ on elements of length $n$ to be in accordance with $f$, as described
above.
Finally, let $M(B_f,y,z)$ denote 1 if $M$ accepts $(x,y,z)$ given oracle
$B_f$, and let
$M(B_f,y,z)$ denote 0 otherwise.
We claim that the following inequality holds for every
$y\in\Sigma^{q(m)}$ and $z\in\Sigma^{r(m)}$:
\begin{equation}
\left|\left\{g\in\mathcal{F}_0\,|\,M(B_g,y,z)=1\right\}\right|
\:\geq\:\left(p-t(m)^2\right)\left|\left\{f\in\mathcal{F}_1\,|\,M(B_f,y,z)=1
\right\}\right|.
\label{eq:accept_count}
\end{equation}
The proof of this inequality is the main technical part of the proof of
Theorem~\ref{theorem:oracle}, and so we postpone this part
momentarily---for now assume that it is proved.

Suppose now that for every $f\in\mathcal{F}_1$ there exists a certificate
$y\in\Sigma^{q(m)}$ such that $M(B_f,y,z) = 1$ for every $z\in\Sigma^{r(m)}$
(which must be the case if $M$ is really a valid machine for solving the Group
Non-Membership problem with respect to an arbitrary oracle).
Since there are only $2^{q(m)}$ possible certificates, we conclude that one of
the certificates must work for many different oracles, i.e., there exists some
fixed $y$ such that for at least $2^{-q(m)}|\mathcal{F}_1|$ choices of
$f\in\mathcal{F}_1$ we have $M(B_f,y,z) = 1$ for every $z\in\Sigma^{r(m)}$.
This implies
\[
\sum_{z\in\Sigma^{r(m)}}\left|\left\{f\in\mathcal{F}_1|M(B_f,y,z)=1\right\}
\right|
\:\geq\: 2^{-q(m)}\,|\mathcal{F}_1|\,2^{r(m)}.
\]
By (\ref{eq:accept_count}) we therefore have
\begin{eqnarray*}
\sum_{g\in\mathcal{F}_0}
\left|\left\{\left.z\in\Sigma^{r(m)}\,\right|\,M(B_g,y,z)=1\right\}\right|
& = & \sum_{z\in\Sigma^{r(m)}}\left|\left\{g\in\mathcal{F}_0\,|\,M(B_g,y,z)=1
\right\}\right|\\
& \geq & (p-t(m)^2)\,2^{-q(m)}\,|\mathcal{F}_1|\,2^{r(m)}.
\end{eqnarray*}
Therefore, there must exist $g\in\mathcal{F}_0$ such that
\begin{eqnarray*}
\left|\left\{\left.z\in\Sigma^{r(m)}\,\right|\,M(B_g,y,z)=1\right\}
\right|
& \geq & \frac{(p-t(m)^2)\,2^{-q(m)}\,|\mathcal{F}_1|\,2^{r(m)}}
{|\mathcal{F}_0|}\\
& > &  2^{-2q(m)}2^{r(m)}.
\end{eqnarray*}

From this we conclude that for any polynomial time oracle Turing machine $M$
and group oracle $B$, there exists an integer $n$ such that by modifying $B$
only on elements of length $n$ it is possible to make $M$ an invalid machine
for the GNM problem; either there exists $f\in\mathcal{F}_1(n)$ such that
no certificate causes $M$ to accept $(1,2)_n$ given group oracle $B_f$ with
certainty, or there exists $g\in\mathcal{F}_0(n)$ such that some certificate
causes $M$ to accept $(1,2)_n$ given group oracle $B_g$ with too high a
probability.

Now it is routine to prove there exists $B$ as in the statement of the theorem
by a diagonalization argument.
Let $(M_1,q_1,r_1),\:(M_2,q_2,r_2),\:\ldots$, be an enumeration of all
triples consisting of a polynomial-time deterministic oracle Turing machine
and a pair of strictly increasing polynomials.
Let $t_1,\:t_2,\:\ldots$ be a sequence of polynomials such that $M_i$ runs
in time $t_i(|x|)$ on each input $(x,y,z)$ and any group oracle $B$, assuming
$|y|=q_i(|x|)$ and $|z|=r_i(|x|)$, for each $i$.
Without loss of generality we may assume $t_{i+1}(m) > t_i(m)$ for all $i$ and
$m$.
We define $B$ using a stage construction as follows:

\begin{list}{}{\setlength{\leftmargin}{0.2in}\setlength{\rightmargin}{0.0cm}
	\setlength{\labelsep}{0.2in}\setlength{\labelwidth}{0in}
	\setlength{\itemindent}{0in}
	\setlength{\itemsep}{0cm}}
\item[{\makebox[0mm][l]{Stage 0:}}] $\;$

Set $B^{(0)}$ to be an arbitrarily chosen group oracle, and set
$n_0=4$.

\item[{\makebox[0mm][l]{Stage $i\geq 1$:}}] $\;$

Choose $n_i$ be the smallest integer satisfying $2n_i > t_{i-1}(4n_{i-1})$ and
$8t_i(4n_i)^2<2^{n_i/2}$, and let $m_i$ be the length of the encoding of the
pair $(1,2)_{n_i}$.

If there exists $f\in\mathcal{F}_1(n_i)$ such that for all
$y\in\Sigma^{q_i(m_i)}$ we have 
\[
\left|\left\{\left.z\in\Sigma^{r(m_i)}\,\right|\,M_i(B^{(i-1)}_f,y,z)=1\right\}
\right| \:<\: 2^{r(m_i)}
\]
then let $B^{(i)} = B^{(i-1)}_f$ for any such $f$.
Otherwise, as proved previously, there exists $g\in\mathcal{F}_0(n_i)$
and $y\in\Sigma^{q_i(m_i)}$ such that
\[
\left|\left\{\left.z\in\Sigma^{r(m_i)}\,\right|\,
M_i(B^{(i-1)}_g,y,z)=1\right\}\right| \:>\: 2^{-2q(m_i)}2^{r(m_i)}.
\]
Set $B^{(i)} = B^{(i-1)}_g$ for any such $g$.
\end{list}
Finally, let $B$ be the group oracle that, for each $i$, agrees with $B^{(i)}$
on all queries regarding elements of length less than $n_{i+1}$.
(This group oracle is well-defined, since all changes to the oracle on
stages subsequent to stage $i$ involve only elements of length at least
$n_{i+1}$.)
It is now straightforward to verify that $\op{GNM}(B)\not\in\op{MA}^B$ by the
construction of $B$, since no triple $(M_i,q_i,r_i)$ can be valid according
to Definition~\ref{def:MA}.

It remains to prove the inequality (\ref{eq:accept_count}).
Define an equivalence relation $\sim_{y,z}$ on $\mathcal{F}\times\mathcal{F}$
for each $y\in\Sigma^{q(m)}$ and $z\in\Sigma^{r(m)}$ as follows:
$f\sim_{y,z}g$ if and only if $f$ and $g$ induce identical executions of $M$
for $x = (1,2)_n$, certificate $y$, and random bits $z$ (i.e., on input
$((1,2)_n,y,z)$).

Let $f\in\mathcal{F}_1$, and consider the computation of $M$ on input
$((1,2)_n,y,z)$ given a group oracle specified by $f$ on length $n$ elements.
During this computation, there will be some number $k$ of queries to the
oracle regarding length $n$ elements, which we may express as
\begin{eqnarray}
u_1 \pm v_1 & = & w_1,\nonumber\\
& \vdots & \label{eq:equations}\\
u_k \pm v_k & = & w_k\nonumber
\end{eqnarray}
(that is, the $i$-th query asks for $u_i+v_i$ or $u_i-v_i$, and the answer
given by the oracle is $w_i$).
Let $L$ denote the set $\{u_1,v_1,w_1,\ldots,u_k,v_k,w_k\}$ (i.e., the
distinct length-$n$ labels of group elements that either appear in
a query or a response), and let $l$ denote the size of $L$.
Without loss of generality assume the labels 1 and 2 are in $L$.
The above equations specify a $k\times l$ matrix $A$ with entries in
$\{-1,0,1\}$ in the following straightforward way: the columns of $A$ are
indexed by the labels in the set $L$, and for each $i = 1,\ldots,k$, the
$i$-th row of $A$ only has nonzero entries corresponding to labels $u_i$,
$v_i$, and $w_i$.
In case the $i$th query was $u_i + v_i = w_i$, the entries for the columns
indexed by $u_i$, $v_i$, and $w_i$ will be $1$, $1$, and $-1$, respectively,
and in case the $i$th query was $u_i - v_i = w_i$, the entries will be
$1$, $-1$, and $-1$, respectively.

At this point it will be convenient to view $\mathbb{Z}_p\times\mathbb{Z}_p$
as being the additive group of the field $\mathbb{F} = GF(p^2)$ in order to
easily apply well-known theorems from linear algebra to our analysis.
(Here the specific correspondence between $\mathbb{Z}_p\times\mathbb{Z}_p$
and $\mathbb{F}$ is arbitrary, so long as the additive group
structure is preserved.)
Note that for any $g$ satisfying \mbox{$f\sim_{y,z} g$}, we must have that the
values $g$ assigns to the labels in $L$ form a vector in the nullspace of $A$
(viewing $A$ as a matrix over $\mathbb{F}$).

Let $d$ be the dimension of the nullspace of $A$.
We claim that
\begin{equation}
\left|\left\{g\in\mathcal{F}_0\,|\,f\sim_{y,z}g\right\}\right|
\:\geq\: \left(p-1-\binom{l}{2}\right)
\left(p^{2d-4} - \binom{l}{2}p^{2d-6}\right)(p^2-l)! \label{eq:equiv1}
\end{equation}
and
\begin{equation}
\left|\left\{g\in\mathcal{F}_1\,\left|\,f\sim_{y,z}g\right.\right\}\right|
\leq p^{2d-4}(p^2-l)!.
\label{eq:equiv2}
\end{equation}
This suffices to prove (\ref{eq:accept_count}), since by (\ref{eq:equiv1}) and
(\ref{eq:equiv2}) we determine that for all $f\in\mathcal{F}_1$ we have
\[
\left|\left\{g\in\mathcal{F}_0\,|\,f\sim_{y,z}g\right\}\right|
\:\geq\: (p-t(n)^2) \left|\left\{g\in\mathcal{F}_1\,\left|\,f\sim_{y,z}g
\right.\right\}\right|,
\]
and summing over those equivalence classes for which $M(B_f,y,z)=1$ yields
(\ref{eq:accept_count}).

The inequality (\ref{eq:equiv2}) is immediate since the collection of vectors
in the nullspace of $A$ that assign values $(1,0)$ and $(0,1)$ to the labels
$1$ and $2$, respectively, is a hyperplane of dimension $d-2$, and each
vector in this hyperplane can be extended to yield at most $(p^2 - l)!$
distinct $g\in\mathcal{F}_1$ with $g\sim_{y,z}f$.

To prove (\ref{eq:equiv1}), let us define
\[
H_a\: =\:\{h\in\mathbb{F}^{\,l}\,|\,Ah = 0,\,h[1] = (1,0),\;\mbox{and}
\;h[2]=(a,0)\}
\]
for each $a\in\{2,\ldots,p-1\}$, and define
\[
T \:=\: \{h\in\mathbb{F}^{\,l}\,|\,h[i]\not=h[j]\;\mbox{for}\;i\not=j\}.
\]
We will prove that there are at least $p-1-\binom{l}{2}$ values of $a$
for which $H_a\cap T$ contains at least $p^{2d-4}-\binom{l}{2}p^{2d-6}$
elements.
As each $h\in H_a\cap T$ may be extended to yield $(p^2 - l)!$ distinct
$g\in\mathcal{F}_0$ with $g\sim_{y,z}f$, we will have proved
(\ref{eq:equiv1}).

Suppose $H_a\cap T$ is nonempty for $a\in\{2,\ldots,p-1\}$.
Then of course $H_a$ is nonempty, and is therefore a hyperplane of dimension
$d-2$.
We may also conclude that for each pair $i\not=j\in L$, the intersection of
$H_a$ with the subspace $J_{i,j} = \{h\in\mathbb{F}^{\,l}\,|\,h[i] = h[j]\}$ is
properly contained in $H_a$, and is therefore a hyperplane of dimension at
most $d-3$.
Since $T = \mathbb{F}^{\,l}\,\backslash\left(\bigcup_{i\not=j}J_{i,j}\right)$,
there must therefore be at least $p^{2(d-2)}-\binom{l}{2}p^{2(d-3)}$
elements in $H_a\cap T$ as required.

Thus, it remains to prove that $H_a\cap T$ is nonempty for at least
$p - 1 - \binom{l}{2}$ values of \linebreak
$a\in\{2,\ldots,p-1\}$.
In order to prove this, define a mapping $\varphi_a:\mathbb{Z}_p\times
\mathbb{Z}_p\rightarrow\mathbb{Z}_p\times\mathbb{Z}_p$ for each
$a\in\{2,\ldots,p-1\}$ as $\varphi_a(\alpha,\beta) = (\alpha + a\beta,0)$.
Let $h_f\in\mathbb{F}^{\,l}$ denote the vector corresponding to the values
assigned to the labels in $L$ by $f$, and let $\varphi_a(h_f)$ denote the
vector obtained by applying $\varphi_a$ to each entry of $h_f$ individually.
Following from the fact that each $\varphi_a$ is a homomorphism, we must have
that $\varphi_a(h_f)$ is in the nullspace of $A$, and therefore
$\varphi_a(h_f) \in H_a$.
Write $h_f[i] = (\alpha_i,\beta_i)$ for each $i$, and suppose we have
$\varphi_a(h_f[i])=\varphi_a(h_f[j])$ for some pair $i\not=j$.
Then $\alpha_i + a\beta_i \equiv \alpha_j + a\beta_j\;(\bmod\,p)$, and so
$a(\beta_i-\beta_j) \equiv \alpha_j -\alpha_i\;(\bmod\,p)$.
Since $h_f[i]\not=h_f[j]$ (as $f$ assigns distinct values to each label), it
is impossible that $\beta_i=\beta_j$, and so
$a\equiv(\beta_i-\beta_j)^{-1}(\alpha_j -\alpha_i)\;(\bmod\,p)$.
It follows that there are at most $\binom{l}{2}$ nonzero values of $a$ such
that $\varphi_a(h_f)\not\in H_a\cap T$, which completes the proof.
\qed

Finally, we consider a restricted case of the Group Non-Membership problem
where there are only two input group elements (i.e., $k=1$ in the statement of
the GNM problem).
\begin{center}
\underline{2-Element Group Non-Membership (2-GNM)}\\[2mm]
\begin{tabular}{ll}
Instance: & Group elements $g$ and $h$ in some group $G$.\\
Question: & Is $h$ outside the group generated by $g$ (i.e., is
$h\not\in\langle g\rangle$)?
\end{tabular}
\end{center}
We note that this problem can be solved in BQP for any group oracle $B$
using Shor's algorithm.
\begin{prop}
$\op{2-GNM}(B)\in\op{BQP}^B$ for any group oracle $B$.
\end{prop}
As this problem is not contained in (classical) MA relative to the group
oracle $B$ constructed in the proof of Theorem~\ref{theorem:oracle}, we
have obtained the relation $\op{BQP}^B\not\subseteq\op{MA}^B$.

\begin{cor}
There exists an oracle $B$ such that $\op{BQP}^B\not\subseteq\op{MA}^B$.
\end{cor}


\section{Other problems having succinct quantum proofs}
\label{sec:other}

\noindent
Quantum certificates for group non-membership may be used in conjunction
with classical certificates for other group properties to obtain succinct
quantum certificates for various problems regarding finite groups.
A few examples are given in this section.

Consider the following problems:

\begin{center}
\underline{Proper Subgroup}\\[2mm]
\begin{tabular}{ll}
Instance: & Elements $g_1,\,\ldots,\,g_k$ and $h_1,\,\ldots,\,h_l$ in
some group $G$.\\
Question: & Is $\,\langle h_1,\:\ldots,\:h_l\rangle\,$ a proper subgroup of
$\langle g_1,\ldots,g_k\rangle$?
\end{tabular}
\end{center}

\begin{center}
\underline{Divisor of Order}\\[2mm]
\begin{tabular}{ll}
Instance: & Elements $g_1,\ldots,g_k$ in some group $G$ and an integer
$N$.\\
Question: & Does $N$ divide the order of $\langle g_1,\ldots,g_k\rangle$?
\end{tabular}
\end{center}

\begin{center}
\underline{Simple Group}\\[2mm]
\begin{tabular}{ll}
Instance: & Elements $g_1,\ldots,g_k$ in some group $G$.\\
Question: & Is $\langle g_1,\ldots,g_k\rangle$ a simple group?
\end{tabular}
\end{center}

\begin{center}
\underline{Intersection}\\[2mm]
\begin{tabular}{@{}ll}
Instance: & Elements $g_1,\:\ldots,\:g_k$, $h_1,\:\ldots,\:h_l$, and
$a_1,\ldots,a_t$ in some group $G$.\\
Question: & Is $\langle a_1,\ldots,a_t\rangle$ equal to the intersection
of $\langle g_1,\ldots,g_k\rangle$ and $\langle h_1,\ldots,h_l\rangle$?
\end{tabular}
\end{center}

\begin{center}
\underline{Centralizer}\\[2mm]
\begin{tabular}{ll}
Instance: & Elements $g_1,\ldots,g_k$, $h_1,\ldots,h_l$ and $a$ in some
group $G$.\\
Question: & Is $\langle h_1,\ldots,h_l\rangle$ equal to the centralizer of
$a$ in $\langle g_1,\ldots,g_k\rangle$?
\end{tabular}
\end{center}

\begin{center}
\underline{Maximal Normal Subgroup}\\[2mm]
\begin{tabular}{ll}
Instance: & Elements $g_1,\,\ldots,\,g_k$ and $h_1,\,\ldots,\,h_l$ in some
group $G$.\\
Question: & Is $\langle h_1,\ldots,h_l\rangle$ a maximal normal subgroup of
$\langle g_1,\ldots,g_k\rangle$?
\end{tabular}
\end{center}
\vspace{1mm}

The first two problems, Proper Subgroup and Divisor of Order, are in
$\op{QMA}^B$ for any group oracle $B$, while neither is in $\op{MA}^B$ for
appropriate choice of $B$.
Quantum certificates for these problems may be obtained by combining quantum
certificates for non-membership with classical certificates for other
properties.

In the case of Proper Subgroup this is straightforward: a quantum proof that
$\langle h_1,\ldots,h_l\rangle$ is properly contained in
$\langle g_1,\ldots,g_k\rangle$ may consist of a classical portion that
certifies that each $h_i$ may be generated from $g_1,\ldots,g_k$ and
identifies an element $a\in\langle g_1,\ldots,g_k\rangle$ that purportedly
lies outside of $\langle h_1,\ldots,h_l\rangle$, while the quantum
portion certifies that $a\not\in\langle h_1,\ldots,h_l\rangle$.

In the case of Divisor of Order, the quantum proof is slightly more
complicated: for each prime power $p^l$ dividing $N$, the quantum proof
identifies a tower of $p$-subgroups
\[
\langle h_1\rangle \:\leq\: \langle h_1,h_2\rangle \:\leq\: \cdots\:\leq\:
\langle h_1,\ldots,h_l\rangle
\]
of $\langle g_1,\ldots,g_k\rangle$ having the property
$h_i\not\in\langle h_1,\ldots,h_{i-1}\rangle$ for each $i$ (so that
$\langle h_1,\ldots,h_l\rangle$ has order at least $p^l$).
The $p$-subgroup property may be certified classically \cite{BabaiS84},
while each $h_i\not\in\langle h_1,\ldots,h_{i-1}\rangle$ may be certified
with a quantum proof of non-membership.

The remaining four problems, Simple Group, Intersection, Centralizer,
and Maximal Normal Subgroup, are in $\op{co-QMA}^B$ for any group oracle $B$.
For the complements of each of these problems, quantum proofs may be
obtained from quantum proofs for non-membership along with classical proofs
for various properties as above.
For the case of Simple Group and Maximal Normal Subgroup, we rely on the
fact that there exist classical certificates for the property of one group
being normal in another \cite{Babai92}.
We leave the details for the reader.


\section{Open Problems}
\label{sec:conclusion}

\noindent
We conclude by mentioning some open problems relating to quantum proofs and
the class QMA.

\begin{itemize}
\item Is Graph Non-Isomorphism in QMA?

\item Is Group Order in QMA?
(That is, given group elements $g_1,\ldots,g_k$ and an integer $N$, are there
succinct quantum proofs for the property $N=|\langle g_1,\ldots,g_k\rangle|$?)

\item Is co-NP contained in QMA?
Do unexpected consequences result from such a containment?

\item We have claimed that $\op{QMA}\subseteq\op{PP}$; can a better
upper-bound be placed on the power of QMA?
What other relations among QMA and other classes can be proved?

\end{itemize}



\bibliographystyle{plain}


\end{document}